\documentclass[
    ,final            % use final for the camera ready runs
  ]
  {aipproc}
\layoutstyle{6x9}
\usepackage{amsmath}
\usepackage{amssymb}
\def\corr{\rm corr}
\def\x{\bf x}

\usepackage{color}
\usepackage{graphicx}
\begin{document}
\title{Electrode Selection for Noninvasive Fetal Electrocardiogram Extraction using Mutual Information Criteria\footnote{\textcolor{blue}{This article is an extended version of the following: R.~Sameni, F.~Vrins, F.~Parmentier, C.~Herail, V.~Vigneron, M.~Verleysen, C.~Jutten, and M.B.~Shamsollahi. ``Electrode selection for noninvasive fetal electrocardiogram extraction using mutual information criteria.'' In \textit{AIP Conference Proceedings}, vol. 872, no. 1, pp. 97-104. American Institute of Physics, 2006. DOI: 10.1063/1.2423265.}}}
	%\texttt{http://www.aip..org/pacs/index.html}>}
\author{Reza Sameni}{
  address={Laboratoire des Images et des Signaux (LIS) -- CNRS UMR 5083, INPG, UJF, Grenoble, France}
  ,altaddress={Biomedical Signal and Image Processing Laboratory (BiSIPL), School of Electrical Engineering, Sharif University of Technology, Tehran, Iran}
}
\author{Fr\'ed\'eric Vrins}{
  address={Machine Learning Group (MLG), Microelectronics Laboratory, Universit\'e Catholique de Louvain (UCL), Louvain-La-Neuve, Belgium}
}
\author{Fabienne Parmentier}{
  address={Machine Learning Group (MLG), Microelectronics Laboratory, Universit\'e Catholique de Louvain (UCL), Louvain-La-Neuve, Belgium}
}
\author{Christophe H\'erail}{
  address={Laboratoire Syst\'emes Complexes (LSC) -- CNRS FRE 2494, Evry, France}
}
\author{Vincent Vigneron}{
  address={Laboratoire Syst\'emes Complexes (LSC) -- CNRS FRE 2494, Evry, France}
}
\author{Michel Verleysen}{
  address={Machine Learning Group (MLG), Microelectronics Laboratory, Universit\'e Catholique de Louvain (UCL), Louvain-La-Neuve, Belgium}
}
\author{Christian Jutten}{
  address={Laboratoire des Images et des Signaux (LIS) -- CNRS UMR 5083, INPG, UJF, Grenoble, France}
}
\author{Mohammad B. Shamsollahi}{
  address={Biomedical Signal and Image Processing Laboratory (BiSIPL), School of Electrical Engineering, Sharif University of Technology, Tehran, Iran}
}
%\begin{center}
%December 2006
%\end{center}

%//////////////////////////////////////////////////////////////////////////////////////////////
\begin{abstract}
Blind source separation (BSS) techniques have revealed to be promising approaches for, among other, biomedical signal processing applications. Specifically, for the noninvasive extraction of fetal cardiac signals from maternal abdominal recordings, where conventional filtering schemes have failed to extract the complete fetal ECG components. From previous studies, it is now believed that a carefully selected array of electrodes well-placed over the abdomen of a pregnant woman contains the required `information' for BSS, to extract the complete fetal components. Based on this idea, in previous works array recording systems and sensor selection strategies based on the Mutual Information (MI) criterion have been developed. In this paper the previous works have been extended, by considering the 3-dimensional aspects of the cardiac electrical activity. The proposed method has been tested on simulated and real maternal abdominal recordings. The results show that the new sensor selection strategy together with the MI criterion, can be effectively used to select the channels containing the most `information' concerning the fetal ECG components from an array of 72 recordings. The method is hence believed to be useful for the selection of the most informative channels in online applications, considering the different fetal positions and movements.
%\textcolor{red}{[TO BE COMPLETED]}
\end{abstract}
\maketitle
%//////////////////////////////////////////////////////////////////////////////////////////////
\section{Introduction}
In recent years, \textit{blind source separation} (BSS) techniques have found special interests in the biomedical signal processing community. Specifically BSS has been very effective for the noninvasive extraction of fetal cardiac signals from maternal abdominal recordings \cite{LMV00}, where conventional filtering schemes do not have satisfactory. From the related state-of-the-art, it is believed that the complete shape of the fetal ECG should be extractable from a sufficient number of recordings captured by electrodes suitably positioned over the abdomen of a pregnant woman. Based on this idea, in previous research multi-channel recording systems containing up to 72 electrodes have been developed, which can be placed as a belt of electrodes over the abdomen and the back of a pregnant woman \cite{VVJV04}. However, the fetal cardiac signals are not directly available from such recordings, for two reasons. First, many of the recording channels are contaminated with the high-power maternal ECG noise and contain little information about the fetal ECG. Second, the processing of all the different combinations of these electrodes (72$\times$71/2 electrode pairs), can be very time-consuming and inefficient, since a much smaller subset of the electrodes (which can also vary with time depending on the location of the fetus, shape of the abdomen, or stage of pregnancy), may be sufficient to extract the required `information'. Based on these facts, an electrode selection strategy was proposed in a recent study to reject the channels which correspond to the maternal ECG, by minimizing the \textit{mutual information} (MI) between the different electrodes and a reference channel reflecting the `pure' maternal ECG \cite{VJV04}.

On the other hand according to the \textit{dipole theory} of the cardiac electrical activity, it is known that the electrodes placed on the body surface receive projections of the cardiac dipole vectors, depending on their relative positions with respect to the maternal and fetal hearts. This suggests that the `pure' maternal and fetal signals are not unique for all electrodes. Consequently the recordings obtained from each of the abdominal electrodes should be compared with their own corresponding reference, rather than a single reference that is shared by all the sensors.

In this work, the problem of electrode selection has been revisited, by considering the 3-dimensional shapes of the cardiac dipole. For this, a 3-dimensional model of the cardiac dipole vector has been used. Using this model the reference channel has been customized for each of the recording channels and the MI of each channel has been calculated with respect to its own reference. Moreover the channel selection strategy has been modified by preserving the channels having the lowest MI with the associated maternal ECGs \textit{and}, at the same time, the highest MI with the fetal ECG (to avoid selecting highly noisy signals). But obviously, the fetal reference ECG is not exactly available. To circumvent this problem, a pre-processing stage is proposed, which consists of detecting the fetal R-peaks in a well-chosen abdominal channel and using it to construct artificial fetal
reference signals.

The proposed method has been tested on both simulated and real recordings, by first removing the baseline wander noises, and then applying the channel selection algorithm. The selected channels have been later decomposed into independent components (ICs) by using the JADE ICA algorithm. The results show that the ICs extracted from 6 to 12 selected channels, very well correspond to the fetal ECG components with the least interference from the
maternal ECG, which is an interesting improvement over the previous sensor selection approach. This shows that the MI criterion together with an appropriate model of the cardiac dipole vector can be used as an effective means of channel selection for the application of interest.
%//////////////////////////////////////////////////////////////////////////////////////////////
\section{Background}
\subsection{The Vectorcardiogram vs. the Electrocardiogram}
\label{sec:VCGvsECG}
The electrical activity of the cardiac muscle and its relationship with the body surface potentials, namely the \textit{Electrocardiogram} (ECG), has been studied with different approaches ranging from \textit{single dipole models} (SDM), to \textit{activation maps} \cite{Dos00}. Among these methods, the simplest and yet the most popular is the SDM, which is believed to explain 80\%--90\% of the representation power of the body surface potentials \cite{Oos02}. The ECG and the vectorcardiogram (VCG) are also based on this model. According to the SDM, the cardiac electrical activity may be represented by a time-varying rotating vector, with its origin located at the center of the heart and its end sweeping a quasi-periodic region in the
space. This vector may be mathematically represented in the Cartesian coordinates, as follows:
\begin{equation}\label{eq:Cartesien}
  \begin{array}{l}
	   \textbf{d}(t)=x(t)\hat{\textbf{i}}+y(t)\hat{\textbf{j}}+z(t)\hat{\textbf{k}},\\
  \end{array}
\end{equation}
where $\hat{\textbf{i}}$, $\hat{\textbf{j}}$, and $\hat{\textbf{k}}$ are the unit vectors of the Cartesian coordinates.

The dipole model is a means of modeling the heart source, and in order to analyze the electrical recordings on the body surface, an additional model is required for the propagation of the heart potentials in the body \textit{volume conductor}. By assuming this volume conductor as a passive resistive medium which only attenuates the source field (no delays, no reverberations, etc), any ECG signal would simply be a linear projection of $\textbf{d}(t)$, onto the direction of the recording electrode axes $\textbf{v}=a\hat{\textbf{i}} + b\hat{\textbf{j}} + c\hat{\textbf{k}}$:
\begin{equation}\label{eq:ECG}
	   ECG(t)=<\textbf{d}(t),\textbf{v}>=a\cdot x(t) + b\cdot y(t) + c\cdot z(t),
\end{equation}

\begin{figure}[tb]
\centering
\includegraphics[width=4in]{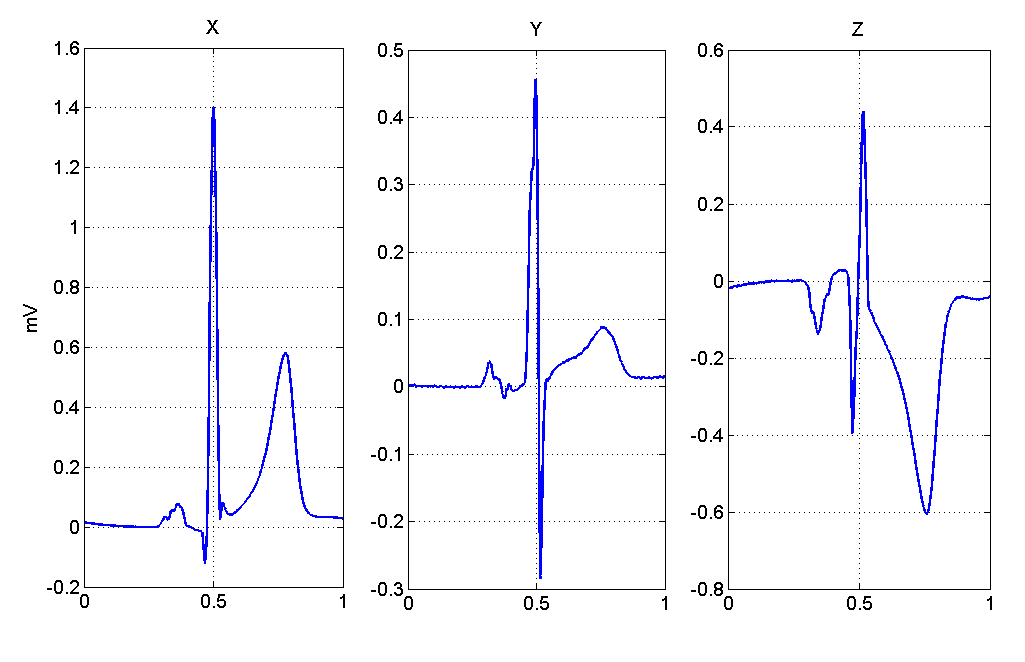}
\caption{Typical Frank lead recordings}
\label{fig:FrankLeads}
\end{figure}

% Fred, I liked the 'triplet linearly independent ...' expression that you used, but 'triplet' has also the meaning of 'three fetus' in this context, and we have used it later in the paper.
A 3-D vector representation of the ECG, namely the VCG, is also possible by using three of such ECG signals. Basically any three linearly independent set of ECG electrodes can be used to construct the VCG. However in order to achieve an orthonormal representation which best resembles the dipole vector $\textbf{d}(t)$, a set of three orthogonal electrodes which best match to the three body axes are selected. The normality of the representation is further achieved by attenuating the different leads with \textit{a priori} knowledge of the body volume conductor, to compensate for the non-homogeneity of the body thorax. The \textit{Frank lead system}, or the \textit{corrected Frank lead system} which have better orthogonality and normalization, are conventional means of recording the VCG \cite{MP95}. Typical signals recorded from the Frank lead electrodes are depicted in Fig. \ref{fig:FrankLeads}.

An ECG recorded from the body surface is contaminated with different noises such as muscle artifacts, electrode/body movements, and baseline wanders \cite{friesen90}, which cause translations of the \textit{isoelectric point} of the cardiac signals in the multi-dimensional VCG space, together with perturbations, rotations, and scalings of the VCG loop.
%//////////////////////////////////////////////////////////////////////////////////////////////
\subsection{The Recording System}
\label{sec:RecSys}
\begin{figure}[tb]
\centering
\includegraphics[width=3.5in]{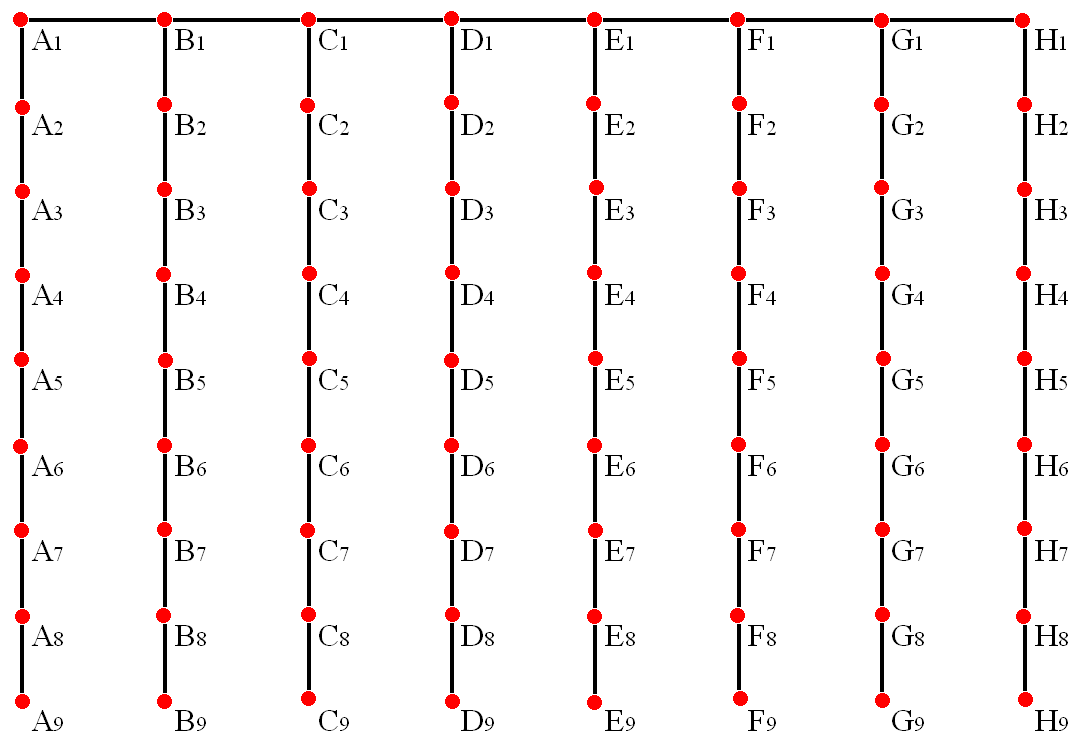}
\caption{The electrode array configuration}
\label{fig:ElectrodeConfig}
\end{figure}

The recording system consists of 72 electrodes placed in a grid of 8 columns and 9 rows throughout the maternal abdomen and back, as illustrated in Fig. \ref{fig:ElectrodeConfig}. The signals recorded from each of the 2 neighboring electrode pairs are amplified by a differential amplifier and sampled at 1\,kHz with a resolution of 12-bits. Accordingly 71 differential channels are achieved, which are transferred to a computer for further processing. Note that, from this electrode configuration, it is possible to reconstruct the potential difference between any pair of the original recording electrodes, through a linear combination of the recorded signals; meaning that a total combination of (72$\times$71)/2 = 2,256 differential pair of recordings is possible.
Apparently, this array configuration can be very rich of information, but at the same time, computationally inefficient. For instance, the proximity of adjacent electrodes implies a high correlation between the associated recordings. In fact there are some previous works which have achieved in the extraction of fetal ECG components by using just between 12--18 electrodes \cite{TST03}. In this work, however, the idea behind using this rather dense grid of electrodes is to be able to have a canonical representation of the fetal ECG regardless of the fetal position and movements. This would allow the further diagnosis procedure to be robust to the location of both the fetus and the sensors. Therefore, it would be interesting to make an online selection of the most informative pairs of electrodes, and possibly switch between the electrodes as the fetus moves. For this we require an effective and robust procedure of channel selection.
%\newline
%\textcolor{red}{- Further explanations or corrections + Fig. \ref{fig:ElectrodeConfig} [For VINCENT and CHRISTOPHE]}\\
%//////////////////////////////////////////////////////////////////////////////////////////////
%\subsection{Interpretation of independent components extracted from the ECG}
\subsection{Application of Independent Components Analysis and Mutual Information for ECG Signals}
\label{sec:InterpICAECG}
%The simplest form of the \textit{Independent Component Analysis} (ICA) problem consists of the estimation of independent sources $s_i(t)$($i=1,...N$), from their blind linear mixtures:
%\begin{equation}\label{eq:ICA}
%	   \textbf{x}(t)=A\cdot \textbf{s}(t)
%\end{equation}
%where $\textbf{x}(t)=[x_1(t),...,x_n(t)]^T$, $\textbf{s}(t)=[s_1(t),...,s_n(t)]^T$, and $A$ is the mixing matrix.

%Mathematically the ICA problem of (\ref{eq:ICA}) almost always gives a solution for the most independent sources.
Mathematically \textit{Independent Component Analysis} (ICA) always gives a solution for the most independent sources contributing to a linear mixture of signals. However when applying ICA to multivariate ECG recordings, the interpretation of the extracted sources and the extraction of meaningful clinical measures from these components need closer study of the ECG and the VCG. In some previous studies, it has been shown that by using the dipole model of the heart (either the SDM or the multi-dipole model), the body surface potentials can be linearly related to the cardiac dipoles \cite{LMV00,Oos02,SJS06}. In \cite{SJS06}, it is further shown that with appropriate baseline wander removal, the columns of the mixing matrix extracted by ICA correspond to the most representative axes of the VCG loop scatter, which are closely related to the directions of the main planes in the VCG space. In terms of the ECG they also correspond to the main waves of ECG signals, meaning that the shapes and time delays between the extracted ICs convey information about the movements of the cardiac dipole vector as it rotates in space.

Multi-dimensional ICA, or \textit{Independent Subspace Analysis} (ISA) is an extension of ICA used for the separation of independent subspaces, rather than independent components \cite{Car98}. The privilege of ISA compared with ICA is that it can separate multiple groups of sources that can be dependent within each group but independent from the sources of distinct groups (like the subspace of the fetal ECG vs. the maternal ECG subspace). ISA has been proved as a promising tool for the extraction of fetal cardiac signals from maternal abdominal recordings, provided that the baseline wander of the ECG recordings are removed through pre-processing and the ICs are extracted with respect to the isoelectric point of the heart \cite{SJS06}.
%For this study we shall use a median filter for the baseline wander removal, the isoelectric line extraction method developed in \cite{SJS06}, together with the JADE ICA algorithm (\cite{Cardoso}) for the extraction of the maternal and fetal independent subspaces.

There are several means of measuring the independence between random variables \cite{HKO01}, however the \textit{Mutual Information} (MI) has interesting properties that are specifically suitable for the problem of interest \cite{KSG04}. Considering two random vectors (or scalars) of $X$ and $Y$, with the joint probability density functions of $p_{X,Y}(x,y)$, and the marginal densities of $p_{X}(x)$ and $p_{Y}(y)$, the MI is defined as follows:
\begin{equation}\label{eq:MI}
	\displaystyle I(X,Y) = \int \int p_{X,Y}(x,y) log \frac{p_{X,Y}(x,y)}{p_{X}(x)p_{Y}(y)}dx dy
\end{equation}
One of the interesting properties of MI, is the independence with respect to re-parameterizations \cite{KSG04}. This means that if $X'=F(X)$ and $Y'=G(Y)$ are smooth and invertible transformations, we have:
\begin{equation}\label{eq:MIproperty}
	\displaystyle I(X',Y') = I(X,Y)
\end{equation}
These equations hold for both scalar and vector variables of arbitrary dimensions \cite{KSG04}. We can use this feature to study the sensitivity of the MI criterion to the electrode locations for the problem of interest.

According to the SDM and by assuming a linear propagation medium, up to 80\%--90\% of the representation power of the body surface potentials can be achieved through a linear combination of these signals. This implies that with a good approximation, the VCG representations constructed from any 3 linearly independent reference ECG leads can be linearly mapped to the standard orthonormal VCG representation, as follows:
\begin{equation}\label{eq:bodyECG}
    \displaystyle VCG(t) \approx A \cdot VCG_{0}(t)
\end{equation}
and the approximation results from assumptions of SDM that are not exactly fulfilled in practice. In this model the transformation matrix of $A$ depends on the location of the electrodes and the attenuation of the body volume conductor. The well-known \textit{Dower transformation} and its inverse which are used to map the standard ECG leads and the Frank leads, are the evident results of this approximation \cite{EP88}.
Now by using (\ref{eq:MI}) and (\ref{eq:MIproperty}), if we calculate the MI between any noisy recording of $s(t)$ and any set of reference VCGs (not necessarily orthogonal, nor orthonormal), we find the following result:
\begin{equation}\label{eq:MIforBodyECG}
    \displaystyle I(s(t),VCG(t)) = I(s(t),B\cdot VCG_{0}(t)) = I(s(t),VCG_{0}(t))
\end{equation}
where $B=A^{-1}$. Moreover according to (\ref{eq:MIproperty}) the transformation between $VCG(t)$ and $VCG_{0}(t)$ only needs to be invertible and not necessarily linear. So the interesting interpretation of this equation is that as long as a unique and invertible transformation (either linear or nonlinear), exists between two sets of VCG recordings, the MI of any of the recording channels is robust to the changes (and movements) of the reference VCG channels. %Of course, this property may practically be used only for the maternal ECG, as we do not have the three dimensional references for the fetus. 
%\newline
%\textcolor{red}{- Definition of Mutual Information and why we are using it.... [For FREDERIC]}\\
%//////////////////////////////////////////////////////////////////////////////////////////////
\section{A Synthetic Fetal ECG Generator}
\label{sec:dipmodel}
In a recent work \cite{SJCJ06}, a realistic model of the adult ECG and maternal abdominal recordings has been developed. This model is based on the SDM, and consists of the following 3-dimensional dynamic model for the maternal and fetal cardiac dipoles:
\begin{equation}\label{eq:SyntheticDipole}
\begin{array}{l}
	   \displaystyle\dot{\theta}=\omega\\
	   \displaystyle\dot{x}=-\sum_{i}\frac{\alpha^x_i\omega}{(b^x_i)^2}\Delta\theta^x_i exp[-\frac{(\Delta\theta^x_i)^2}{2(b^x_i)^2}]\\
	   \displaystyle\dot{y}=-\sum_{i}\frac{\alpha^y_i\omega}{(b^y_i)^2}\Delta\theta^y_i exp[-\frac{(\Delta\theta^y_i)^2}{2(b^y_i)^2}]\\
	   \displaystyle\dot{z}=-\sum_{i}\frac{\alpha^z_i\omega}{(b^z_i)^2}\Delta\theta^z_i exp[-\frac{(\Delta\theta^z_i)^2}{2(b^z_i)^2}]
\end{array}
\end{equation}
where $\Delta\theta^x_i=(\theta-\theta^x_i)mod(2\pi)$, $\Delta\theta^y_i=(\theta-\theta^y_i)mod(2\pi)$, $\Delta\theta^z_i=(\theta-\theta^z_i)mod(2\pi)$, and $\omega=2\pi f$, where $f$ is the reciprocal of the beat-to-beat heart rate (either for the mother or the fetus). In this model, each of the three coordinates of the dipole vector $\textbf{d}(t)$ is modeled by a summation of Gaussian functions with the amplitudes of $\alpha^x_i$, $\alpha^y_i$, and $\alpha^z_i$; widths of $b^x_i$, $b^y_i$, and $b^z_i$; and located at the rotational angles of $\theta^x_i$, $\theta^y_i$, and $\theta^z_i$. Each of the model parameters in (\ref{eq:SyntheticDipole}) are considered as random variables with appropriate deviations.

As suggested in \cite{SJCJ06}, this model of the cardiac dipole vector can be further related to body surface ECG recordings as follows:
\begin{equation}\label{eq:RecordedECGvsDip}
	ECG(t) = H\cdot R \cdot \Lambda \cdot s(t) + W(t)
\end{equation}
where $ECG(t)_{N\times 1}$ is a vector of the ECG channels recorded from $N$ leads, $s(t)_{3\times 1}=[x(t),y(t),z(t)]^T$ contains the three components of the dipole vector $\textbf{d}(t)$, $H_{N\times3}$ corresponds to the body volume conductor model, $\Lambda_{3\times3}=diag(\lambda_x,\lambda_y,\lambda_z)$ is a diagonal matrix corresponding to the scaling of the dipole in each of the $x$, $y$, and $z$ directions, $R_{3\times3}$ is the rotation matrix for the dipole vector, and $W(t)_{N\times 1}$ is the noise in each of the $N$ ECG channels at the time instance of $t$. Note that $H$, $R$, and $\Lambda$ matrices are generally functions of time.

%Although the product of $H\cdot R \cdot \Lambda$ may be assumed to be a single matrix, the representation in (\ref{eq:RecordedECGvsDip}) has the benefit that the rather stationary features of the body volume conductor that depend on the location of the ECG electrodes and the conductivity of the body tissues can be considered in $H$, while the temporal inter-beat movements of the heart can be considered in $\Lambda$ and $R$, meaning that their average values are identity matrices in a long term study: $E_t\{R\}=I$, $E_t\{\Lambda\}=I$. In Appendix \ref{app:GenRotMat}, a means of coupling these matrices with external sources such as the respiration and achieving non-stationary mixtures of the dipole source is presented.

\begin{figure}[tb]
\centering
\includegraphics[width=4in]{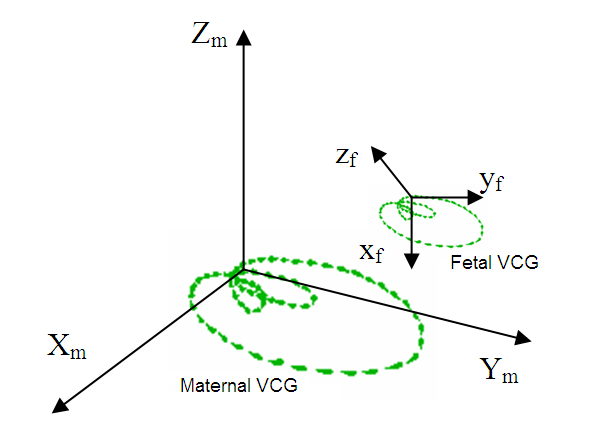}
\caption{Illustration of the fetal and maternal VCGs vs. their body coordinates}
\label{fig:mVCGfVCG}
\end{figure}

Moreover, by utilizing a dynamic model like (\ref{eq:SyntheticDipole}), the signals recorded from the abdomen of a pregnant woman, containing the fetal and maternal heart components can be modeled as follows:
\begin{equation}\label{eq:AbdominalECG}
	X(t) = H_m\cdot R_m \cdot \Lambda_m \cdot s_m(t) + H_f\cdot R_f \cdot \Lambda_f \cdot s_f(t) + W(t)
\end{equation}
where the matrices $H_m$, $H_f$, $R_m$, $R_f$, $\Lambda_m$ and, $\Lambda_f$ have similar definitions as the ones in (\ref{eq:RecordedECGvsDip}), with the subscripts of $m$ and $f$ referring to the mother and the fetus, respectively. In this model, $R_f$ has the additional interpretation that its mean value ($E_t\{R_f\}=R_0$) can be assumed as the relative position of the fetus with respect to the axes of the maternal body. This is an interesting feature for modeling the fetus in the different typical positions such as \textit{Vertex} (fetal head-down) or \textit{Breech} (fetal head-up) positions \cite{WebMD}. As illustrated in Fig. \ref{fig:mVCGfVCG}, $s_f(t)=[x_f(t),y_f(t),z_f(t)]^T$ can be assumed as a canonical representation of the fetal dipole vector which is defined with respect to the fetal body axes, and in order to calculate this vector with respect to the maternal body axes, $s_f(t)$ should be rotated by the 3-dimensional rotation matrix of $R_0$.
%///////////////////////////////////////////////////////////////////
%///////////////////////////////////////////////////////////////////

The model presented in (\ref{eq:AbdominalECG}) may be simply extended to multiple pregnancies (twins, triplets, quadruplets, etc.), by considering independent dynamic models for each of the fetuses.

This model is later used for the generation of synthetic maternal abdominal recordings, by assuming the maternal volume conductor as an infinite homogeneous propagation medium. In this simplified case, the potential generated by the dipole source of $\textbf{d}(t)$, at a distance of $|\textbf{r}|$ is:
\begin{equation}\label{eq:DipolePot}
\displaystyle
\phi(t)-\phi_0= \frac{\textbf{d}(t)\cdot\textbf{r}}{4\pi\sigma|\textbf{r}|^3}= \frac{1}{4\pi\sigma}[x(t)\frac{r_x}{|\textbf{r}|^3} + y(t)\frac{r_y}{|\textbf{r}|^3} + z(t)\frac{r_z}{|\textbf{r}|^3}]
\end{equation}
where $\phi_0$ is the reference potential, $\textbf{r}=r_x\hat{\textbf{a}}_x+r_y\hat{\textbf{a}}_y+r_z\hat{\textbf{a}}_z$ is the vector which connects the center of the dipole to the observation point, and $\sigma$ is the conductivity of the volume conductor \cite{Ges89,MP95}. Now consider the fact that the ECG signals recorded from the body surface are the potential differences between two different points. Equation (\ref{eq:DipolePot}) therefore indicates how the coefficients $a$, $b$, and $c$ in (\ref{eq:ECG}) can be related to the radial distance of the electrodes and the volume conductor material. Of course, in reality the volume conductor is neither homogeneous nor infinite, leading to a much more complex relationship between the dipole source and the body surface potentials.
%//////////////////////////////////////////////////////////////////////////////////////////////
\section{Electrode Selection Strategy}
\label{sec:ElecSelecStrat}
The electrode selection strategy originally suggested in \cite{VJV04}, uses a unique reference signal for the rejection of the channels which contain the most mutual information with the maternal ECG. However according to the 3D representation of the VCG presented in previous sections, it is now clear that the different electrodes are recording the projection of the cardiac potentials from different angles. Hence, although they are all time-synchronized, the maternal ECG contribution in the recordings may have totally different shapes depending on the associated sensor's location. This means that a unique maternal reference template is not sufficient for all channels. Ideally, we would like to have an appropriate reference channel for each of the electrodes. Unfortunately the idea of using separate references for each electrode can only be achieved on simulated data, and it is not applicable to real ECG recordings. However, a suitable compromise is to use a set of 3 reference recordings such as the 3 signals captured from the standard Frank lead electrodes.%Indeed, as stated before,
%\textcolor{red}{[TO BE COMPLETED]}
%[\textcolor{red}{Problem} in the above subsection we have to find the results about electrode selection + ICA based on the simulated pure maternal ECG reference ! Otherwise, there is no comparison between simulation and real-world experiments. The results that are presented therein are not related at all to the ones given below !]

As explained above, there are totally 71 channels of real recordings available. These channels are generally mixtures of different source signals coming from distinct physical sources and noises. In practice, it happens that some of the electrodes are detached from the maternal body, making there corresponding channel totally noisy. These electrodes should be eliminated before the main processing, leaving $N\leq71$ channels for the processing. Among the remaining electrodes, in order to select say $K\ll N$ sensor signals, we shall use the same algorithm as in \cite{VJV04} up to two main differences. First, contrarily to the previous method, the set ${\cal R}$ of reference signals contains more than one component $n({\cal R})>1$. Second, the selection of a $k$-th signal, $1\leq k\leq K$, involves the mutual information with a \textit{local} reference $r(x_k)\in{\cal R}$, chosen according to the candidate $x_k$ in the set $\cal X$ of the recordings.

\begin{figure}[tb]
\centering
\begin{minipage}{3.5in}
\includegraphics[width=4in]{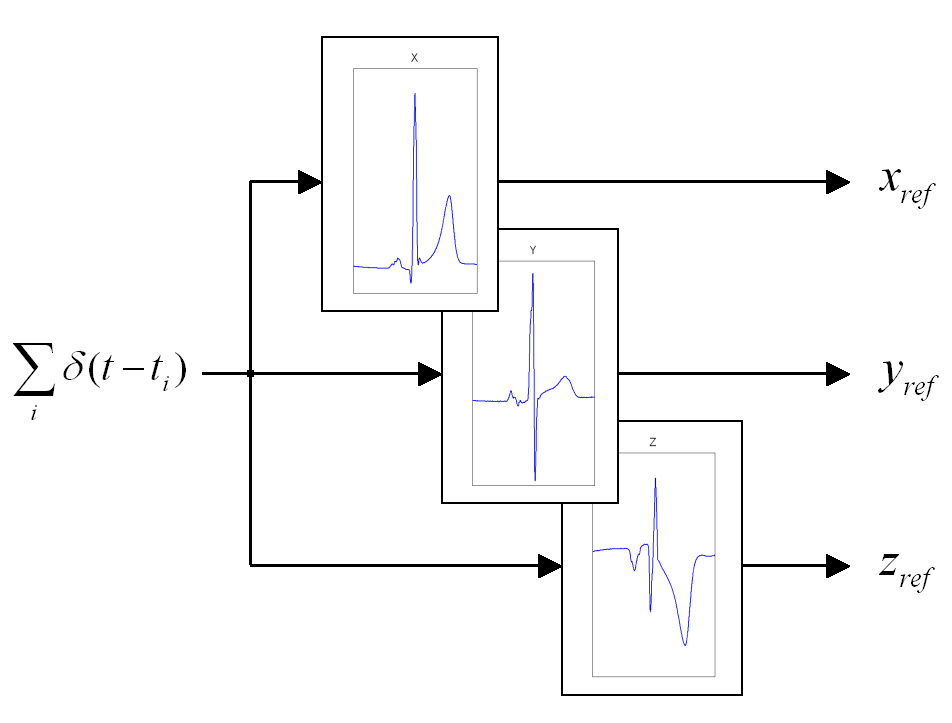}
\end{minipage}
\caption{Illustration of the procedure of generating synthetic Frank lead reference signals from a train of pulses time-synchronized with the ECG R-peaks ($t_i$)}
\label{fig:TrainOfPulses}
\end{figure}

For the simulated data by considering a grid of 144 electrodes, we will study the case of $n({\cal R})=144$, with one fetal and one maternal reference channel for each recorded channel. However, as stated before, for real recordings the `pure' reference signal is not available for each channel. Hence for real recordings, in order to select the channels with the least interference from the maternal components, two experiments have been carried out. In the first experiment we select $n({\cal R})=3$, with the set ${\cal R}=\{r_1(t),r_2(t),r_3(t)\}$ being the set of three typical signals obtained via the Frank lead sensors from a pregnant woman. 
%In this case we restrict our study to $n({\cal R})=3$ and the set ${\cal R}=\{r_1(t),r_2(t),r_3(t)\}$ is chosen to be the set of the three typical signals obtained via the Frank lead sensors from a pregnant woman.
For the case where the Frank lead recordings are not valid, synthetic Frank lead references can be generated by passing a train of pulses which are time-synchronized with the R-peaks of the mother, through systems with impulse responses similar to the typical signals of the Frank lead electrodes (Fig. \ref{fig:TrainOfPulses}). The output of this system can be used as the three reference channels for the maternal ECG.
Using these reference signals, each of the $N$ channels are classified to one of the three classes labelled $1,2,3$, as follows:
%the case here). Obviously, these signals have to be first processed in order to be, among other, synchronized with the recordings. Then, the recordings are classified in three classes, labeled $1,2,3$ (see Figure \ref{fig:CorrMax}), as follows:
\begin{equation}
%\forall x\in{\cal X},~{\cal C}(x)\doteq{\mathop{argmax}_{j\in\{1,\ldots,n({\cal R})\}}}|{\corr}(x,r_j)|,~r_j\in{\cal R}
\forall x\in{\cal X},~{\cal C}(x)\doteq{\mathop{argmax}_{j\in\{1,2,3\}}}|{\corr}(x,r_j)|,~r_j\in{\cal R}
\label{eq:ClassifRule}
\end{equation}
Accordingly each channel is assigned to the class with which it has the most `similarity' with its reference signal, and for further calculations the MI is calculated between each channel and its corresponding class reference.

This pre-classification is used in the following iterative channel selection. Assuming that ${\cal X}'=\{x'_1,\ldots, x'_{k-1}\}$ is the set of $k-1$ already selected signals, the $k$-th signal $x'_k$ is selected as follows:
\begin{eqnarray}
x'_k\doteq \mathop{argmin}_{x_j\in{\cal X}\setminus {\cal X}'} \Big\{I(x_j,r_{{\cal C} (x)}) + \sum_{i=1}^{k-1}I(x'_i,x_j)\Big\},\hspace{.2cm}x'_i\in{\cal X}',r_{{\cal C} (x)}\in{\cal R}\nonumber\hspace{.3cm}\hbox{(\textit{Maternal~rule~1})}
\end{eqnarray}
The intuition behind the first term in this equation is to find the most independent channels (in terms of MI), which contain the least information about the maternal components, while the second term assures that the selected channels are not redundant and contain information which was not provided by the previously selected channels.

In a second experiment, we artificially extend the number of references from $n({\cal R})=3$ to $n({\cal R})=N$ as follows. According to the SDM explained in previous sections, if only the maternal heart was active, the signals recorded from the maternal body could be approximated by:
\begin{equation}
{\x}(t)\simeq H\times VCG(t),
\end{equation}
where ${\x}(t)\in\mathbb{R}^{N}$ is the array of recorded signals, $VCG(t)\doteq[r_1(t),r_2(t),r_3(t)]^T$ is the 3D maternal VCG that can be obtained via the Frank orthonormal representation, and $H\in\mathbb{R}^{N\times 3}$ is the transfer matrix depending on the propagation medium and sensor locations, which may be found by solving the following least squares problem:
\begin{equation}
	\hat{H} \doteq \mathop{argmin}_{H}\|{\x}(t) - H\times VCG(t)\|
\end{equation}

Note that if we consider $\hat h_i$ as the $i$-th row of $\hat H$, $\hat h_i\times VCG(t)$ is the local maternal ECG reference for the $x_i$ recorded signal. This result leads us towards the second sensor selection rule. Again, assuming that ${\cal X}'=\{x'_1,\ldots, x'_{k-1}\}$ is the set of $k-1$ selected signals, the $k$-th signal $x'_k$ is selected as follows:
\begin{eqnarray}
x'_k\doteq \mathop{argmin}_{x_j\in{\cal X}\setminus {\cal X}'} \Big\{I(x_j,\hat h_i\times VCG(t)) + \sum_{i=1}^{k-1}I(x'_i,x_j)\Big\},\hspace{.5cm}x'_i\in{\cal X}'\nonumber\hspace{.5cm}\hbox{(\textit{Maternal~rule~2})}
\end{eqnarray}

Using either of the rules (1) or (2), the channels containing the least MI with the maternal signals are achieved. However this is not enough to ensure that a channel contains fetal components, since highly noisy signals which neither correspond to the maternal nor the fetal components can also be selected by such criteria. In order to resolve this problem, we need some additional information concerning the fetal components. For the moment, lets suppose that we had one or more fetal references defined by the set ${\cal R'}$, similar to the maternal references considered above. In this case, the signal which contains the most information about the fetal components could be found as follows:
\begin{eqnarray}
x''\doteq \mathop{argmax}_{x'_j\in {\cal X}'} \Big\{I(x'_j,r_{{\cal C} (x'_j)})\Big\},\hspace{.5cm}r'_{{\cal C} (x'_j)}\in{\cal R'}\nonumber\hspace{.5cm}\hbox{(\textit{Fetal~rule~1})}
\end{eqnarray}
This rule may be applied several times in order to find the best $M<n({\cal X}')$ channels containing the most information about the fetus, by each time removing the selected channel from ${\cal X}'$ and applying the rule to its remaining members. Note that in the fetal rule, we no longer considered the minimization of the mutual information with the previously selected channels, since the fetal components are very weak and even minor information concerning the fetal components should not be lost.

Now the remaining problem is how to find a set of fetal reference potentials. As mentioned before, for simulated recordings we have the fetal and maternal references for each channel; however for real recordings no \textit{a priori} reference is available for the fetus. In order to find a fetal reference we can apply ICA to the total $N$-channel database or to the channels selected by the maternal rules. This usually provides at least one channel which corresponds to the fetal components and may be used as a single fetal reference. Moreover the R-peaks of the fetal components may also be detected from this single component and used to make a synchronous averaging of the different channels. In fact due to the quasi-periodic shape of the ECG, when we average the different channels synchronous with the fetal R-peaks, the SNR of the fetal components are improved; leading to an average ECG waveform which can be used as the reference for each channel. Next, as illustrated before in Fig. \ref{fig:TrainOfPulses}, we can use these average waveforms together with the R-peak pulses to achieve synthetic references for each channel\footnote{The idea of synchronous averaging may also be used as an alternative means of achieving the maternal references for each channel. However we have not considered this approach in this paper.}. Note however that this approach of fetal reference selection is computationally inefficient, and only helps us evaluate the performance of the proposed method in offline applications.
%//////////////////////////////////////////////////////////////////////////////////////////////
\section{Results}
\label{sec:Results}
\subsection{Simulated Abdominal Signals}
\begin{figure}[tb]
\centering
\includegraphics[width=4in]{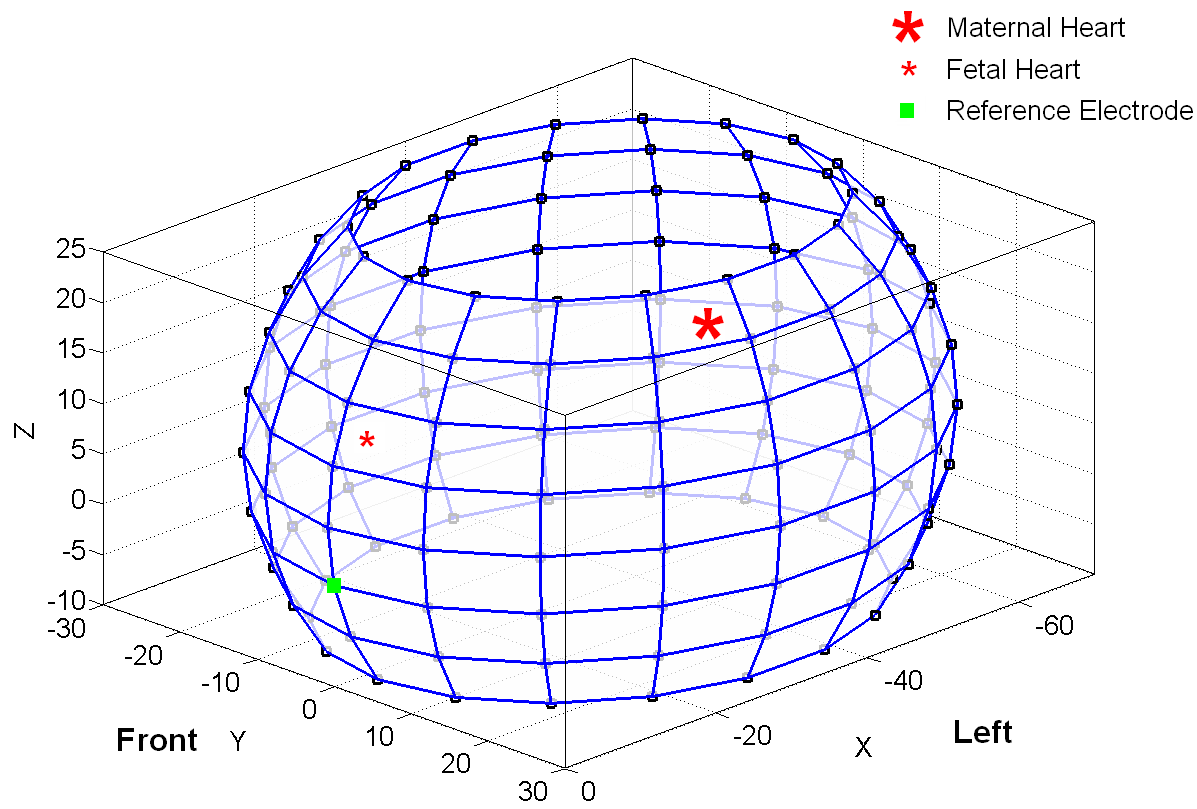}
\caption{The simulated electrode belt scheme with the locations of the maternal and fetal hearts}
\label{fig:ElecConfig}
\end{figure}

As our first study we shall consider a simulated electrode belt containing 144 electrodes placed in an 8$\times$18 grid around the maternal abdomen, as shown in Fig. \ref{fig:ElecConfig}. This electrode configuration is twice as dense as the real system and enables us to study the effect of electrode positions with a higher resolution. In this configuration similar to the electrode configuration suggested in \cite{TST03}, the maternal navel has been assumed as the reference of potentials. Moreover we have assumed the fetus in \textit{Breech} (fetal head-up) position with the fetus faced towards the left hand of the mother. This information is used to calculate the fetal rotation matrix in (\ref{eq:AbdominalECG}). The locations of the assumed maternal and fetal hearts are also indicated in Fig. \ref{fig:ElecConfig}, with the maternal heart close to the upper-left electrodes and the fetal heart close to the lower-right electrodes.
%///////////////////////////////////////////////////////////////////
%///////////////////////////////////////////////////////////////////

\begin{figure}[tb]
\centering
\begin{minipage}{5.5in}
\includegraphics[width=2.75in]{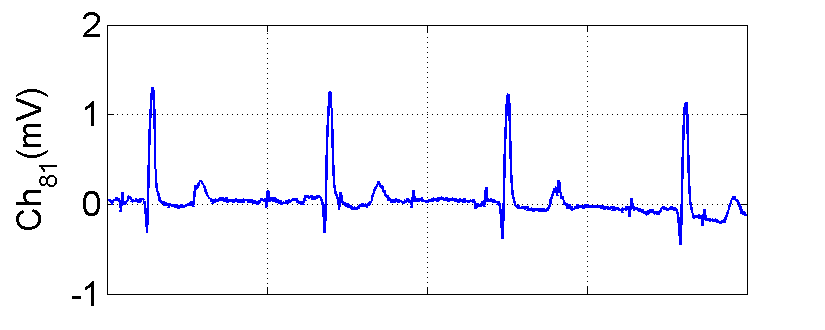}\includegraphics[width=2.75in]{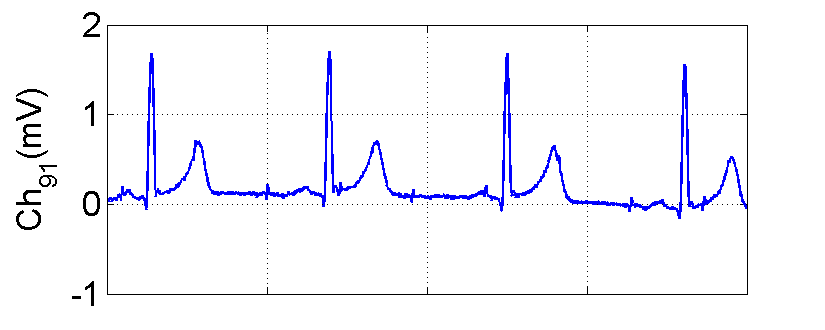}
\includegraphics[width=2.75in]{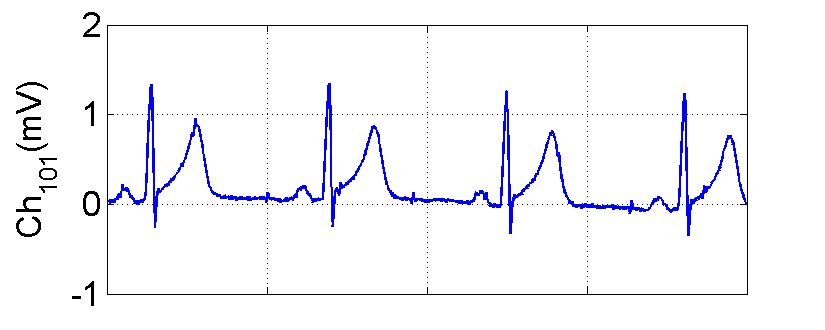}\includegraphics[width=2.75in]{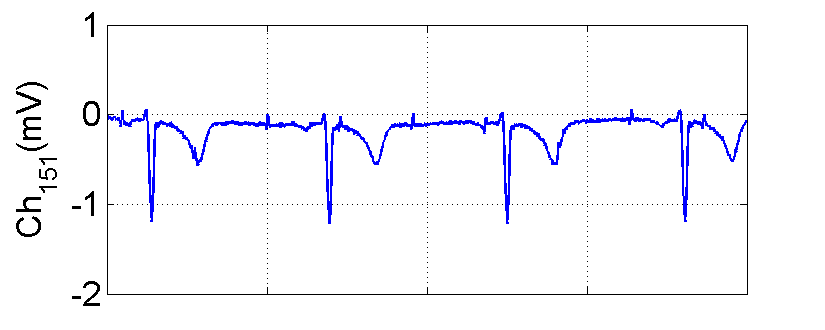}
\end{minipage}
\caption{Typical simulated abdominal signals containing the maternal and fetal ECGs. The small spikes (with a frequency of almost twice the ECGs), correspond to the fetal ECG peaks.}
\label{fig:SampleECGs}
\end{figure}

Now by using the synthetic ECG model presented in section \ref{sec:dipmodel}, and by assuming the maternal volume conductor as a infinite homogeneous medium, we can calculate the true and noise-free maternal contribution to each of the abdominal recordings, without the presence of the fetus. The resultant signals can in fact be considered as the `pure' maternal references for each of the 144 electrodes. The fetal reference signals may also be calculated by only considering the fetal heart in the model. Next we consider both the maternal and fetal hearts, together with some non-stationary noise added to each channel. These simulated signals resemble real ECG recordings made from the maternal abdomen. Here we have assumed the maternal and fetal parameters in (\ref{eq:SyntheticDipole}) to be identical, except that the fetal dipole amplitude (the $\alpha^x_i$, $\alpha^y_i$, and $\alpha^z_i$ parameters) are 1/10$^{th}$ of the maternal dipole amplitudes. The details of the synthetic ECG calculations are similar to those described in \cite{SJCJ06}. Typical samples of these recordings are plotted in Fig. \ref{fig:SampleECGs}.
%///////////////////////////////////////////////////////////////////
%///////////////////////////////////////////////////////////////////

\begin{figure}[tb]
\centering
\begin{minipage}{3.0in}
\includegraphics[width=3.0in]{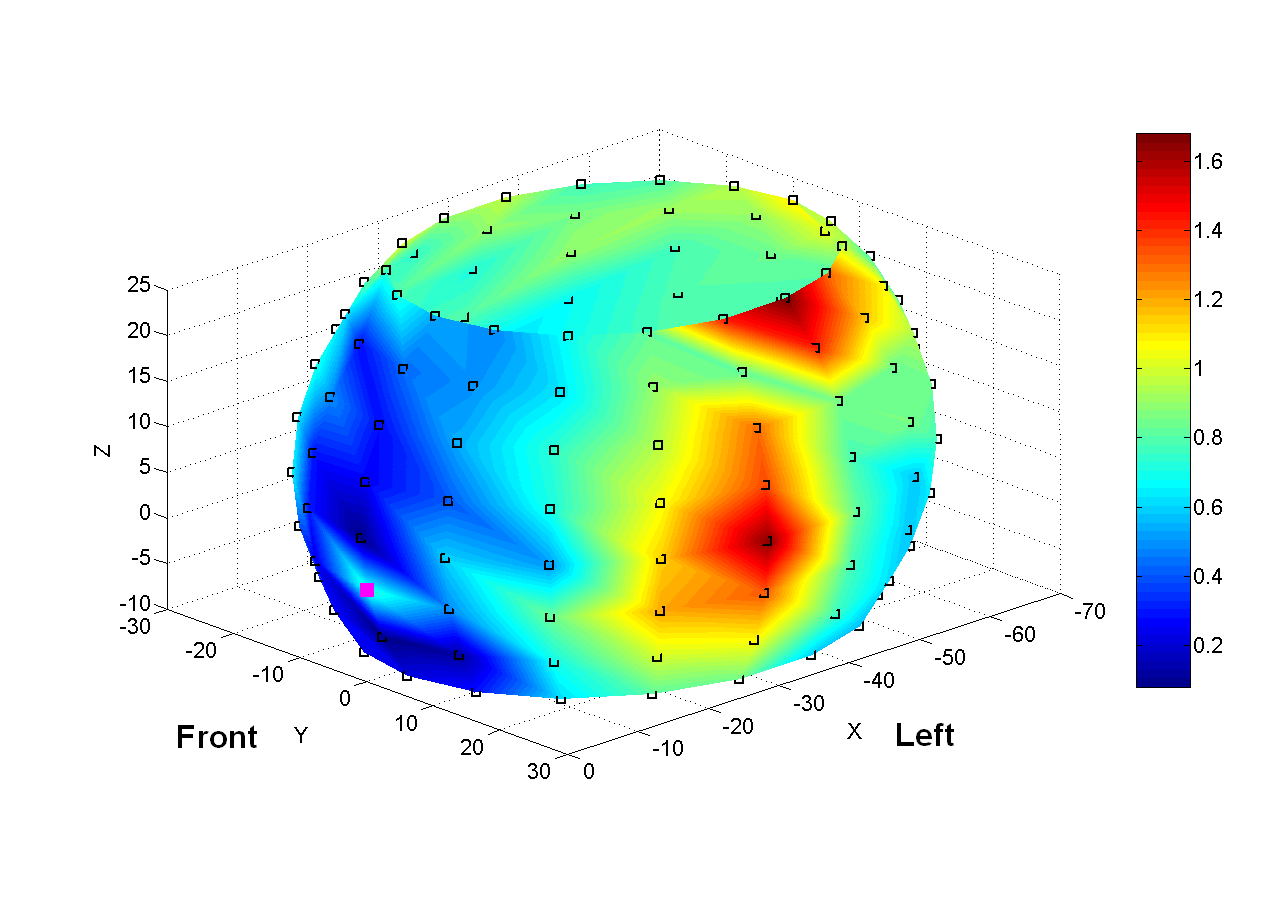}
\begin{center}
\raisebox{.2in}[.25in]{{\footnotesize(a)}}
\end{center}
\end{minipage}
\begin{minipage}{3.0in}
\includegraphics[width=3.0in]{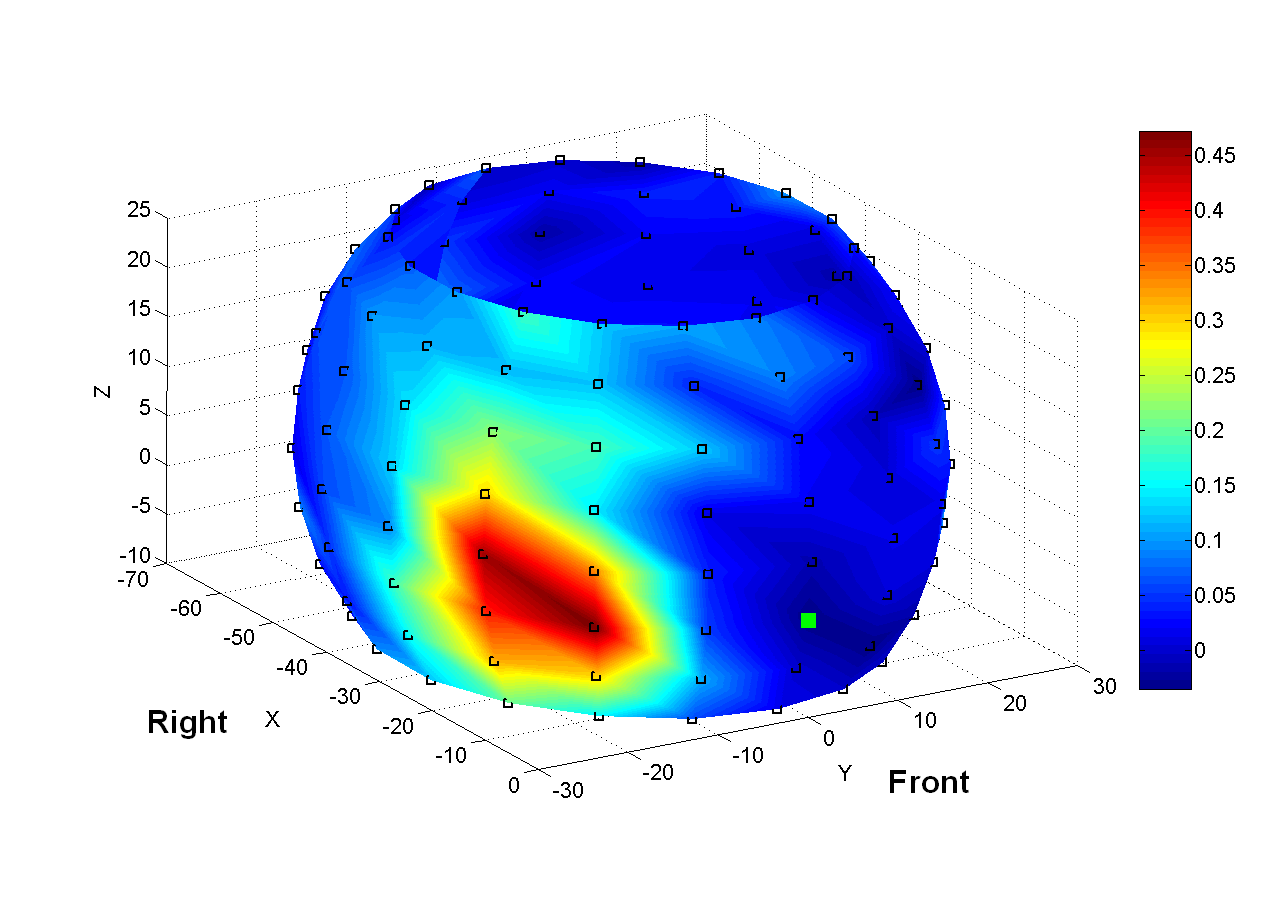}
\begin{center}
\raisebox{.2in}[.25in]{{\footnotesize(b)}}
\end{center}
\end{minipage}
\caption{The MI between the noisy simulated signals and the `true' references for the (a) mother and (b) fetus}
\label{fig:MIPlots}
\end{figure}

We can now calculate the MI between each of the noisy measurements and their corresponding maternal and fetal reference signals. The results of this study are depicted in Fig. \ref{fig:MIPlots}, for both the maternal and fetal reference signals.
%///////////////////////////////////////////////////////////////////
%///////////////////////////////////////////////////////////////////
As it can be seen in Fig. \ref{fig:MIPlots}, the MI criterion shows high values in specific electrodes which are rather close to the heart sources. Specifically, for the maternal heart due to the strong dipole character of the heart the high MI electrodes are not localized in one region, but rather show two distinct regions of high MI.

\begin{figure}[tb]
\centering
\begin{minipage}{3.5in}
\includegraphics[width=3.5in]{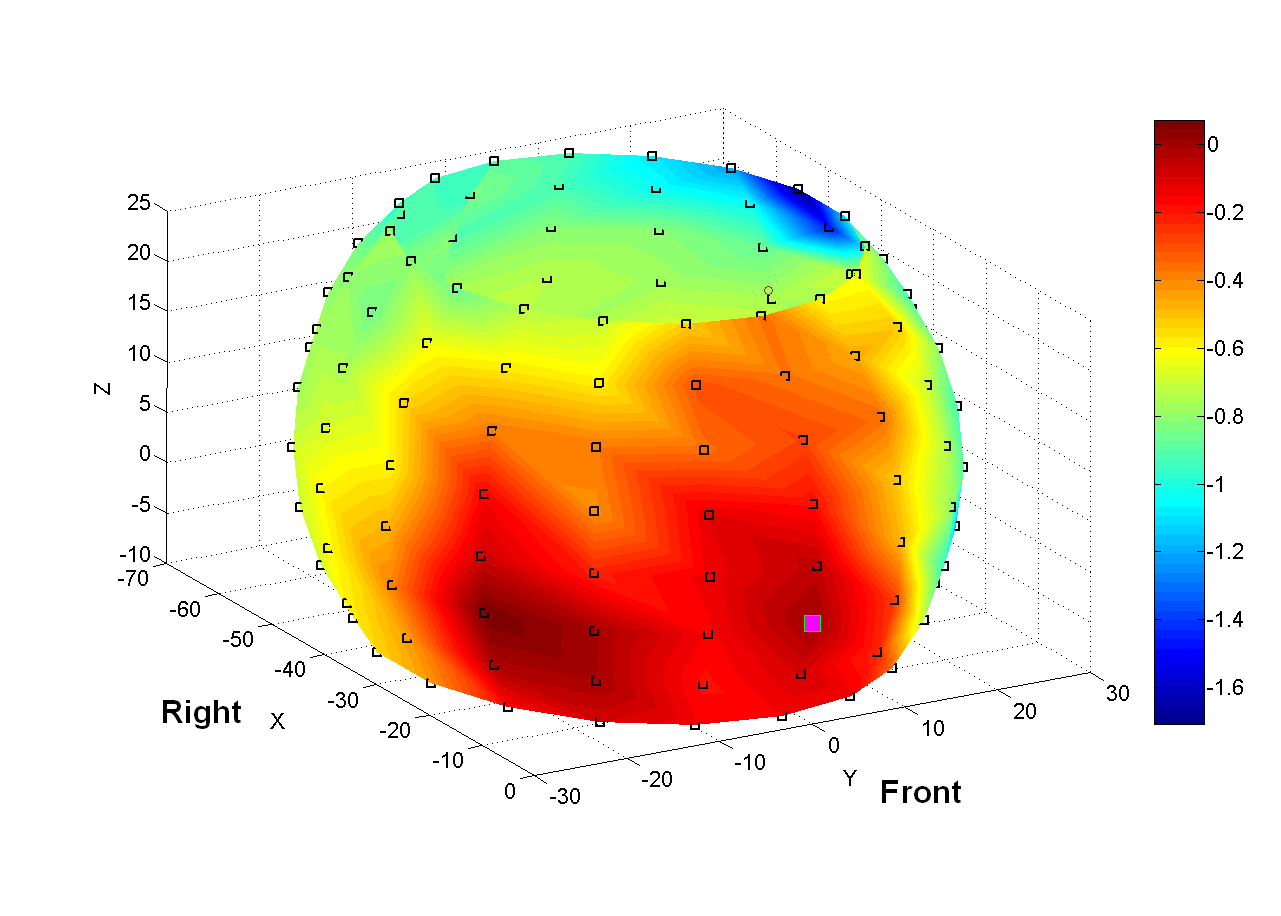}
\end{minipage}
\caption{The MI of each noisy channel with respect to the fetus minus the MI with respect to the mother}
\label{fig:FetalMaternalMI}
\end{figure}

\begin{figure}[tb]
\centering
\begin{minipage}{3.5in}
\includegraphics[width=3.5in]{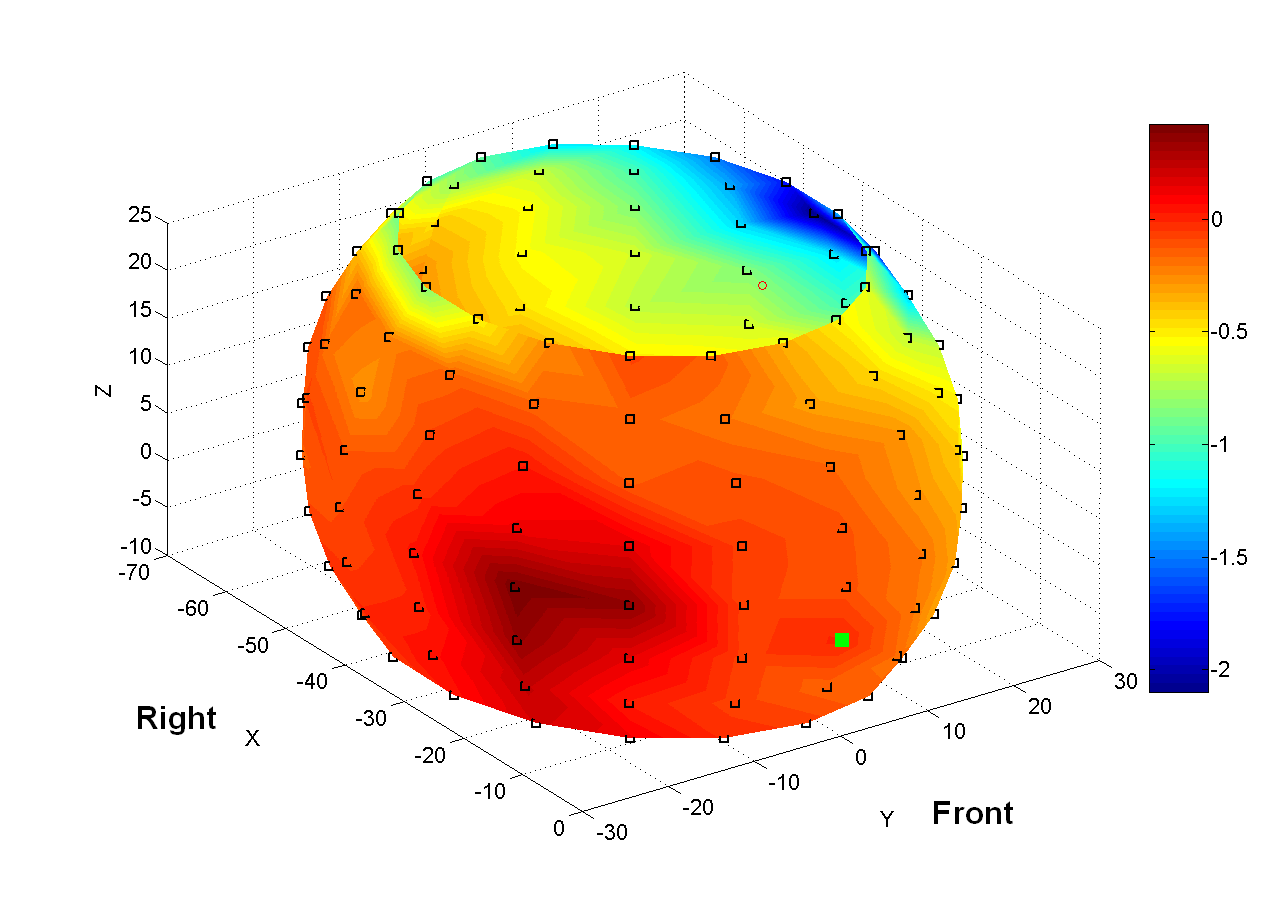}
\end{minipage}
\caption{The MI of each noisy channel with respect to the fetus minus the MI with respect to the mother by using differential recordings between neighboring electrodes.}
\label{fig:FetalMaternalMI2}
\end{figure}

From these results, the difference between the MI calculated from the fetal and maternal references for each electrode, can be considered as a measure for finding the electrodes which contain the maximum information of the fetal ECG and the minimum information of the maternal ECG. This result is plotted in Fig. \ref{fig:FetalMaternalMI}. As seen in this figure, the differential MI criterion clearly shows the electrodes close to the fetal heart and, at the same time, far from the maternal heart as suitable candidates, containing high fetal information. Apparently this region can move according to the fetal position and movements. We can also see in the recent figures that the MI of the electrodes close to the reference electrode (maternal navel) are rather distorted due to the zero potential of this point. This problem may be solved by modifying the reference of potential. For example, we can make differential measurements between adjucent electrodes instead of measuring all the potentials with respect to the same reference. The effect of this change on the results of Fig. \ref{fig:FetalMaternalMI} are plotted in Fig. \ref{fig:FetalMaternalMI2}. As we see the MI is no longer distorted close to the maternal navel.
%///////////////////////////////////////////////////////////////////
%///////////////////////////////////////////////////////////////////
%///////////////////////////////////////////////////////////////////
%//////////////////////////////////////////////////////////////////////////////////////////////
\subsection{Real Recordings}
\begin{figure}[tb]
\begin{minipage}{1.9in}
\centering
\includegraphics[width=1.85in]{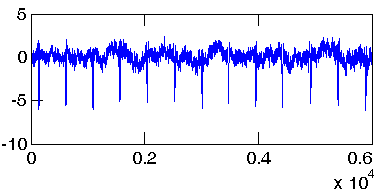}
\begin{center}
\raisebox{.2in}[.1in]{{\footnotesize(a)}}
\end{center}
\end{minipage}
\begin{minipage}{1.9in}
\includegraphics[width=1.9in]{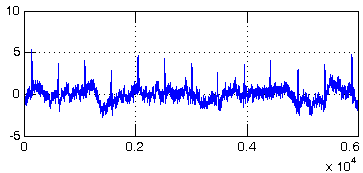}
\begin{center}
\raisebox{.2in}[.1in]{{\footnotesize(b)}}
\end{center}
\end{minipage}
\begin{minipage}{1.9in}
\includegraphics[width=1.9in]{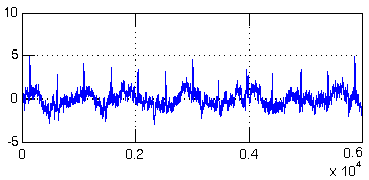}
\begin{center}
\raisebox{.2in}[.1in]{{\footnotesize(c)}}
\end{center}
\end{minipage}
\caption{Typical segment of fetal ECG components extracted by ICA from various sets of sensors using the ICA inputs as (a) $\mathcal{X}$; the 58 signals with relevant information, (b) the 20 signals selected using maternal rule (1) from $\mathcal{X}$, (c) ICA inputs were the 10 signals selected by applying fetal rule (1) from the 20 ones obtained through maternal rule (1).}
\label{fig:ClassifICA}
\end{figure}

%The recording system consists of 72 electrodes placed in a grid of 8 columns and 9 rows throughout the maternal abdomen and back. The signals recorded from each of the 2 neighboring electrode pairs are amplified by a differential amplifier and sampled at 1kHz with a resolution of 12-bits. Accordingly 71 differential channels are achieved, which are transferred to a computer for further processing. Note that, from this electrode configuration, it is possible to reconstruct the potential difference between any pair of the original recording electrodes, through a linear combination of the recorded signals, meaning that a total combination of 72$\times$71/2 differential pair of recordings is possible.
Some of the electrode selection strategies developed in the previous section have been tested on real recordings. The studied database consisted of 71 recordings. Due to the detachments of some electrodes from the skin of the pregnant woman, 13 of these channels were totally noisy channels which were detected and removed through the preprocessing. From the remaining set $\mathcal{X}$ of 58 channels, the maternal R-peaks were detected and the three Frank reference signals were generated by using the R-wave train of pulses. Using these adapted references, these channels were classified into one of the three classes based on equation (\ref{eq:ClassifRule}). In this stage, 47 of the channels were classified to the first class, 6 to the second, and 5 to the third. Next the 20 ones having the least interference from the maternal ECGs were selected by using maternal rule (1). From these 20 channels, 6 were chosen with fetal rule (1); for this stage, the fetal component obtained by running FastICA on the 58 recordings set was chosen as a unique fetal reference\footnote{It should be noted that in this stage, due to the high dimension of the dataset, FastICA was preferred to JADE.}. This `full-set' ICA  has only to be applied once, to build the reference, which can then be used exactly as the Frank leads. Note that if one selects 10 signals from $\mathcal{X}$ using maternal rule (1), JADE does not retrieve any fetal component from the selected recordings; without the fetal rule (1), 20 signals have to be selected by maternal rule (1) in order that JADE gives a fetal component. By contrast, our results indicate that selecting 10 signals using maternal rule (2) from $\mathcal{X}$ is sufficient: a fetal rule is not needed here. Indeed, both methods maternal rule (1) + fetal rule (1), and the maternal rule (2) alone, give globally similar signals. Specifically for $K=10$, 7 sensors are common to both sets. Typical fetal components extracted by the different methods can be seen in Fig. \ref{fig:ClassifICA}. We can see in this figure that the SNR is slightly better in the signal extracted from the total 58 channels; but the fetal components have still been extracted by using a selected subset of 10 channels.
%//////////////////////////////////////////////////////////////////////////////////////////////
\section{Discussions and Conclusions}
\label{sec:Conclusion}
In this paper, the extraction of the fetal ECG from a large array of sensors is investigated in the ICA framework. In order to address this problem in real-time, a selection of few sensors from the array is considered. In order to perform an ICA method on a subset of say $k\ll N$ recordings, an ad-hoc selection strategy has to be developed, since applying ICA on a randomly selected subset of $k$ recordings does not give a fetal component, except when $k$ is rather high (close to $N$). This paper extends previous works on sensor selection strategies. The presented approaches use multidimensional references inspired by a detailed analysis of the cardiac signals and their dimensionality. The proposed strategy is proved on real-world data. Contrary to a random selection technique, our method gives a small subset which contains fetal information. Indeed, in spite of a strong dimensionality reduction of the ICA input space, the fetal component has been retrieved. The method is hence believed to be useful for the selection of the most informative channels in online applications, considering the different fetal positions and movements.

%//////////////////////////////////////////////////////////////////////////////////////////////
%\begin{theacknowledgments}
%The authors would like to thank ....
%\end{theacknowledgments}
%//////////////////////////////////////////////////////////////////////////////////////////////



\bibliographystyle{aipproc}   % if natbib is available

\bibliography{IEEEabrv,References}

%%%%%%%%%%%%%%%%%%\bibliography{sample}
%%%%%%%%%%%%%%%%%%
%%%%%%%%%%%%%%%%%%%%%%%%%%%%%%%%%%%%%%%%%%%%%%%%%%%%%%%%%%%%%
%%%%%%%%%%%%%%%%%%%% Just a reminder that you may have to run bibtex
%%%%%%%%%%%%%%%%%%%% All of it up to \end{document} can be removed
%%%%%%%%%%%%%%%%%%%% if you don't like the warning.
%%%%%%%%%%%%%%%%%%%%%%%%%%%%%%%%%%%%%%%%%%%%%%%%%%%%%%%%%%%%%
%%%%%%%%%%%%%%%%%%\IfFileExists{\jobname.bbl}{}
%%%%%%%%%%%%%%%%%% {\typeout{}
%%%%%%%%%%%%%%%%%%  \typeout{******************************************}
%%%%%%%%%%%%%%%%%%  \typeout{** Please run "bibtex \jobname" to optain}
%%%%%%%%%%%%%%%%%%  \typeout{** the bibliography and then re-run LaTeX}
%%%%%%%%%%%%%%%%%%  \typeout{** twice to fix the references!}
%%%%%%%%%%%%%%%%%%  \typeout{******************************************}
%%%%%%%%%%%%%%%%%%  \typeout{}
%%%%%%%%%%%%%%%%%% }

\end{document}